\documentclass[showpacs,prd,reprint,preprintnumbers,amsmath,amssymb,superscriptaddress,nofootinbib]{revtex4-1}

\usepackage{graphicx,subfigure,multirow}
 \usepackage{longtable}
\usepackage{dcolumn}
\usepackage{bm}
\usepackage[all]{xy}
\usepackage{color}
\usepackage{slashed}
\usepackage[usenames,dvipsnames]{xcolor}
\usepackage[unicode=true,pdfusetitle,
 bookmarks=true,bookmarksnumbered=false,bookmarksopen=false,citecolor=Turquoise,
 breaklinks=false,pdfborder={0 0 1},backref=false,colorlinks=true]
 {hyperref}
 
 \usepackage[normalem]{ulem}

\usepackage{ulem}

\makeatletter

\newcommand{\fmslash}[2][0mu]{%
  \mathchoice
    {\fmsl@sh\displaystyle{#1}{#2}}%
    {\fmsl@sh\textstyle{#1}{#2}}%
    {\fmsl@sh\scriptstyle{#1}{#2}}%
    {\fmsl@sh\scriptscriptstyle{#1}{#2}}}
\newcommand{\fmsl@sh}[3]{%
  \m@th\ooalign{$\hfil#1\mkern#2/\hfil$\crcr$#1#3$}}
\makeatother

\newcommand{\lsim}{{\;\raise0.3ex\hbox{$<$\kern-0.75em\raise-1.1ex\hbox{$\sim$}}\;}}
\newcommand{\gsim}{{\;\raise0.3ex\hbox{$>$\kern-0.75em\raise-1.1ex\hbox{$\sim$}}\;}}

\newcommand{\beq}{\begin{equation}}
\newcommand{\eeq}{\end{equation}}
\newcommand{\bea}{\begin{eqnarray}}
\newcommand{\eea}{\end{eqnarray}}
\mathchardef\minus="002D

\usepackage{color}

\hypersetup{
	colorlinks = true,
	linkcolor = blue,
	citecolor = magenta
}

\begin{document}
\title{Axion-Mediated Inelastic Dark Matter}
\author{Kyu Jung Bae}
\email{kyujung.bae@knu.ac.kr}
\affiliation{
Department of Physics, Kyungpook National University, Daegu 41566, Korea}
\author{Jongkuk Kim} 
\email{jkkim@kias.re.kr}
\affiliation{School of Physics, Korea Institute for Advanced Study, Seoul 02455, Korea}
\affiliation{Department of Physics, Chung-Ang University, Seoul 06974, Korea}

\preprint{KIAS-P23066}

\begin{abstract}
{
We consider the axion-mediated scattering processes between dark matter (DM) and nucleus.
Substantial contributions are made via the CP-odd gluonic current which induces the spin-dependent process.
Since the QCD axion is too feebly coupled to the visible particles, non-QCD axions are necessary 
for the current DM experiments to accomplish the ample sensitivity.
In the case of multi-component DM models, the inelastic scattering processes 
also make sizable contributions to the direct detection.
The supersymmetry (SUSY) and clockwork (CW) mechanism provide a realistic model for the QCD and non-QCD axions and the axion-mediated DM scattering processes. 
In the SUSY CW axion model, the lightest axino is the DM particle and the axions mediate the elastic and inelastic scattering processes.
We show that the current and future XENONnT can produce relevant constraints for some parameter space of the model.
}
\end{abstract}


\maketitle

\section{Introduction}

Dark matter (DM) is one of the most essential ingredients 
which form the current universe.
In the early universe, DM develops the gravitational potential without generating the pressure,
and consequently forms the structures much earlier than the case with only baryons.
Its energy density is accurately measured by observing the cosmic microwave background (CMB)~\cite{Planck:2018vyg},
and large scale structure (LSS)~\cite{eBOSS:2020yzd,DES:2021wwk}.
While these measurements are firmly established via the gravitational interaction,
no other interactions of DM are known and the identity of DM particles remains unanswered.\footnote{For a recent review, we refer the readers to Ref.~\cite{Bertone:2016nfn}.}

For a plausible explanation of the observed DM abundance, 
one may introduce weakly interacting massive particles (WIMPs) 
which are frozen out from the early universe due to their weak scale interactions with the visible particles,
{\it i.~e.}, the standard model (SM) particles~\cite{Jungman:1995df,Bertone:2004pz}. 
Among various ways to introduce weak-scale interactions,
one lucrative scenario is that DM particles have neutral current interactions.
DM species is one of electric neutral components of weak multiplets, so it has couplings of order the weak interaction.
This scenario provides a good framework for the DM abundance, direct detection, and indirect detection without introducing new force carriers~\cite{Goldberg:1983nd,Ellis:1983wd,Dienes:1998vg,Cheng:2002iz,Cirelli:2005uq}.
The Higgs portal is also an intriguing scenario which has a great discovery potential 
in diverse experiments~\cite{Patt:2006fw}.
In the other beyond-the-SM (BSM) models, DM interactions can be mediated by new gauge bosons~\cite{Holdom:1985ag}, and neutrinos~\cite{Falkowski:2009yz}.\footnote{For a phenomenological study on various DM models, we refer the readers to Ref.~\cite{Arcadi:2017kky}.}

In this article, we consider another possibility that the DM scattering is mediated by the {\it axions}.
The axion was originally introduced to solve the strong CP problem.
It is the pseudo-Nambu-Goldstone boson (pNGB) of the broken U(1) Peccei-Quinn symmetry,
and automatically relaxes the QCD $\theta$ parameter to be zero~\cite{Peccei:1977hh, Weinberg:1977ma,Wilczek:1977pj}.
The axion is indeed a good DM candidate. Its decay constant is of order the intermediate scale so that
its lifetime is much larger than the age of the universe and the interactions are highly suppressed~\cite{Kim:1979if,Shifman:1979if,Zhitnitsky:1980tq,Dine:1981rt}.
Its coherent motion near the potential minimum can satisfy the observed DM density~\cite{Abbott:1982af,Preskill:1982cy,Dine:1982ah,Turner:1985si} 
and produces unique features in the detection experiments~\cite{Sikivie:1983ip}.
Since these properties can also originate from the string landscape, a lot more axion-like particles arise in the high energy theories
and they show a wide spectrum with various masses and decay constants~\cite{Arvanitaki:2009fg}. 
From now on, we use the term axion for all axion-like particles and the QCD axion is used only for axions 
whose mass is generated via the QCD confinement.

If axions couple to the DM,
they can mediate the DM-nucleus scattering processes in the direct detection experiments.
The QCD axion may be considered as the simplest possibility because it is the indispensable ingredient for resolving the strong CP problem.
In this case, however, the interaction must be highly suppressed by the intermediate scale decay constant of the QCD axion,
and thus the axion cannot make any significant contribution to the DM direct detection.
In this respect, axions with smaller decay constants ({\it i.~e.}, larger interactions) are required
for substantially contributing to the DM direct detection.

A simple way is 
to introduce the axion with smaller decay constant regardless of its origin.
Although there may be some constraints from the collider experiments~\cite{Mimasu:2014nea, Bauer:2017ris}, flavor probes~\cite{Bauer:2021mvw}, beam dump experiments~\cite{Dobrich:2015jyk} and cosmology/astrophysics~\cite{Depta:2020wmr,Balazs:2022tjl},
one can build a model where the axion mediation makes sizable contributions to the DM direct detection.
In this paper, however, we consider a more systematic model which not only incorporates the sizable axions-DM interactions
but also involves the QCD axion.


The {\it clockwork} mechanism~\cite{Choi:2014rja,Choi:2015fiu,Kaplan:2015fuy} makes it possible to realize the QCD axion with the intermediate scale decay constant
from $N\sim {\cal O}(10)$ pseudoscalar fields with the weak scale decay constants.
The zero mode of the clockwork model becomes the QCD axion whose interactions are exponentially suppressed 
while the heavier modes keep their weak scale interactions~\cite{Higaki:2015jag}.
Once the theory is supersymmetrized, 
it automatically embraces the fermion partners of axions which we call axinos.
The axinos indeed couple to the axions in the clockwork model, and they inherit the weak scale interactions from the axions.
Therefore the lightest axino can be a good DM candidate
and its scattering processes is mediated by a series of clockwork axions~\cite{Bae:2020hys}.

Furthermore, we have the same number of axino states and their mass difference can be small compared to their absolute mass scale.
If this is the case, we can also see the {\it inelastic} scattering processes in the direct detection experiments~\cite{Tucker-Smith:2001myb}.
In the detector, we will see the collective signature of various exited DM states and mediators.

This paper is organized as follows.
In Sec.~\ref{sec:single}, we consider a simple toy model which contain a non-QCD axion and a single-component Majorana fermion DM
to illustrate how the axion-mediated process contributes to the DM-nucleon scattering.
In Sec.~\ref{sec:2-comp}, we consider a 2-component DM model to argue 
the kinematics for the inelastic scattering process.
In Sec.~\ref{sec:susy_axion_model}, we briefly review the SUSY CW axion model.
In Sec.~\ref{sec:susy_axion_result}, we present numerical analyses for the axino-mediated DM scattering 
including both the elastic and inelastic processes in the SUSY CW axion model.
In Sec.~\ref{sec:conclusion}, we conclude this paper.

\section{Axion-Mediated Process and Direct Detection}
\label{sec:single}

In this section, 
we investigate a simple model of an axion mediating the scattering process between the visible and DM sectors.
It is a good for describing the essence of the axion mediation in the DM-nucleus scattering.
The Lagrangian is given by
\begin{eqnarray}
{\cal L}&=&{\cal L}_{\rm kin}+\frac12 m_{\chi}\bar{\chi}\chi - V(a)\nonumber \\
&&+\frac{g_s^2}{32\pi^2 f} aG_{\mu\nu}^b \tilde{G}^{b\mu\nu}
-\frac{\xi}{f} \left(\partial_{\mu} a\right)\bar{\chi}\gamma^5\gamma^{\mu}\chi
\label{eq:toy_model}
\end{eqnarray}
where ${\cal L}_{\rm kin}$ is the Lagrangian for kinetic terms, $\chi$ is the Majorana fermion DM, $m_{\chi}$ is the DM mass, $a$ is the axion, $G^{b}_{\mu\nu}$ and $\tilde{G}^{b}_{\mu\nu}$ are the gluon field strength and its dual,  $g_s$ is the
strong coupling constant, $f$ is the axion decay constant, and $\xi$ is the coupling constant.
For the shift symmetry of the axion, we introduce only the derivative interactions between the axion and DM fields.
We consider the axion-gluon-gluon interaction for simplicity, but there can be the axion-photon-photon interaction.\footnote{For a discussion of the generic axion interactions, see Ref.~\cite{Choi:2021kuy}}
The axion potential is given by
\begin{eqnarray}
V(a)=\Lambda^4 \left[1-\cos\left(\frac{a}{f}\right)\right]
\end{eqnarray}
Here $\Lambda$ is the confinement scale of a QCD-like sector and should be larger than that of QCD, 
so we can avoid the strong constraints that apply to the QCD axion.
The axion mass $m_a$ is thus $\sim\Lambda^2/f$ and can be a free parameter of the model.

\begin{figure}[t]
	\centering
	\includegraphics[width=0.65\linewidth]{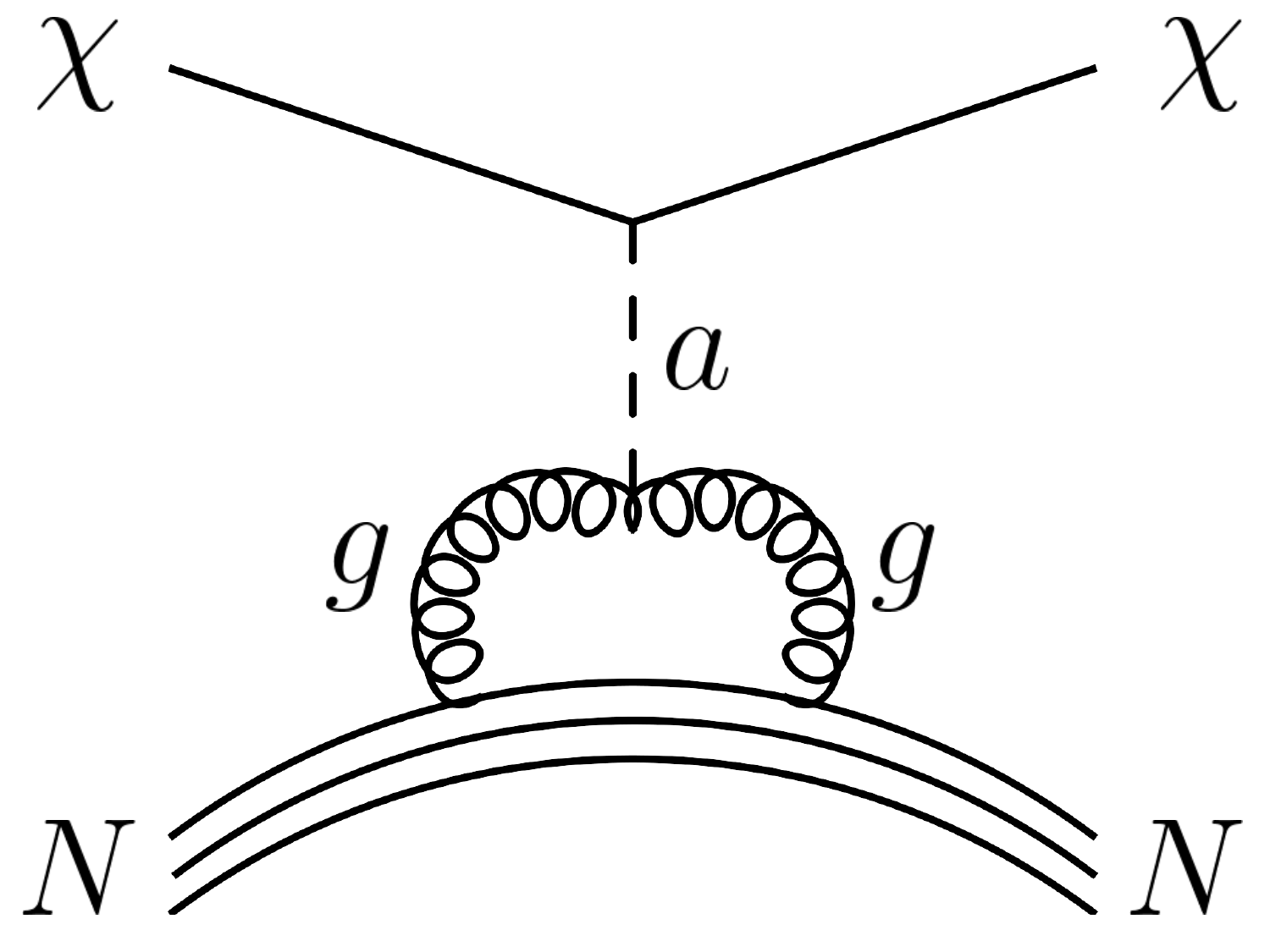} 
	\vspace{0.5cm}
	\caption{ Feynman diagram for elastic scattering process
	} \label{DM-elastic-fig}
\end{figure} 

In this model, the gluon form factor substantially determines the DM-nucleon scattering process
via the CP-odd gluonic current as shown in Fig.~\ref{DM-elastic-fig}.
The matrix element of the CP-odd gluonic current comes from the axion via the QCD chiral anomaly. 
In the large-$N_c$ and chiral limits, the form factor is expressed as~\cite{Bishara:2017nnn}
\begin{eqnarray}
\langle N' \vert \frac{\alpha_s}{8\pi} G^{b\mu\nu}\tilde{G}^b_{\mu\nu} \vert N\rangle &=& F^N_{\tilde{G}}(q^2) \bar{u}'_N i\gamma_5 u_N,
\label{eq:agg_form_factor}
\end{eqnarray}
where $\bar{u}'_N$ is the outgoing nucleon spinor, $u_N$ is the incoming nucleon spinor, and the momentum transfer is expressed by $q^2$. 
Taking into account the leading-order and next-to-leading-order terms, the form factor $F^N_{\tilde{G}}$ is given by
\begin{align}
F^N_{\tilde{G}}(q^2) =& ~{m_N} \biggl( \frac{q^2}{m^2_\pi -q^2} a^{N}_{\tilde{G},\pi} +\frac{q^2}{m^2_\eta -q^2} a^{N}_{\tilde{G},\eta}+b^{N}_{\tilde{G}} +\cdots \biggr), \nonumber\\
\equiv&~ m_Na_N(q^2)
\label{eq:nuc_form_factor}
\end{align}
where $m_N$ is the nucleon mass.
The CP-odd gluonic current contains light-meson poles, and thus the function of $a_N$ depends on momentum transfer, $q^2$.  
The effective mass parameter and coupling constants are given by
\begin{align}
\frac{1}{\tilde{m}} &=\frac{1}{m_u}+\frac{1}{m_d}+\frac{1}{m_s},\\ 
a^{N}_{\tilde{G},\pi} &\approx -\tilde{m} \frac{g_A}{2}\left(\frac{1}{m_u} -\frac{1}{m_d}\right),\\
a^{N}_{\tilde{G},\eta} &\approx -\frac{ \tilde{m} }{6} \left( \Delta u +\Delta d -2\Delta s\right) \left( \frac{1}{m_u} +\frac{1}{m_d} -\frac{2}{m_s}\right), \\
b^{N}_{\tilde{G}} &\approx  -\tilde{m} \left(\frac{\Delta u}{m_u} +\frac{\Delta d}{m_d} +\frac{\Delta s}{m_s}  \right), 
\end{align}
Here $\Delta u = 0.847, \Delta d = -0.407, \Delta s = -0.035,$ and $g_A = \Delta u - \Delta d = 1.254$ \cite{FlavourLatticeAveragingGroupFLAG:2021npn}.
In the case for small momentum transfer, the dominant contribution comes from $b^{N}_{\tilde{G}}$.

The interaction Lagrangian for the DM-nucleon scattering can be written in the following form:
\begin{eqnarray}
{\cal L}={\cal O}_{\chi N} (\bar{\chi}\gamma^5\chi)(\bar{N}\gamma^5N)\, ,
\label{eq:DM-nuc_int}
\end{eqnarray}
with $N=p,n$. 
Here DM part $\bar{\chi}\gamma^5\chi$ is derived from the equation of motion from the Lagrangian in Eq.~\eqref{eq:toy_model},
and the nucleon part $\bar{N}\gamma^5N$ comes from the CP-odd gluonic current in Eq.~\eqref{eq:agg_form_factor}.
The effective operator ${\cal O}_{\chi N}$ is given by
\begin{eqnarray}
{\cal O}_{\chi N}=-\left(\frac{\xi}{f^2}\frac{4m_{\chi}m_N}{q^2-m_a^2}\right)a_N(q^2)\, .
\end{eqnarray}
Notice that the momentum transfer in DM-nucleus scattering is $|q^2|\sim (100{\rm MeV})^2 $ for the DM mass of ${\cal O}(100)$~GeV.
For the heavy axions ($m_a^2\gg |q^2|$), the $q^2$-dependence in ${\cal O}_{\chi N}$ arises only from the nucleon form factor in Eq.~\eqref{eq:nuc_form_factor},
and the effective theory approach is appropriate.
For the light axions ($m_a^2\ll q^2$), on the other hand, ${\cal O}_{\chi N}$ is not an effective operator since ${\cal O}_{\chi N}\sim q^{-2}$,
but it is properly incorporated in the non-relativistic limit, $q^2\to 0$.

In the non-relativistic limit, the four-fermion interaction in Eq.~\eqref{eq:DM-nuc_int} becomes the $q^2$-dependent spin interaction:
\begin{eqnarray}
(\bar{\chi}\gamma^5\chi)(\bar{N}\gamma^5N)\to -4(\vec{S}_{\chi}\cdot \vec{q})(\vec{S}_{N}\cdot \vec{q})\, ,\label{eq:eff_sd}
\end{eqnarray}
where $\vec{S}_{\chi , N}$ is the spin of nucleon and DM particle and $\vec{q}$ is the 3-vector of the momentum transfer.
This corresponds to the ${\cal O}_6$ in the non-relativistic matching of the DM-nucleon scattering in Ref.~\cite{Fitzpatrick:2012ix}.
The spin-dependent DM elastic scattering process was computed in the previous literature. 
In Ref.~\cite{Arina:2014yna}, the only effective operators are considered, so it provides the bound for the heavy mediator cases.
In Ref. \cite{Dolan:2014ska}, the light mediator effects are also included and the LUX bounds are considered.
In this work, we properly derive the XENONnT bounds including the heavy and light mediator cases.

The spin-averaged matrix element for the DM-nucleus scattering is thus given by
\begin{eqnarray}
\overline{\left| {\cal M}_T\right|^2}
&=&\frac14 \left(\frac{\xi^2}{f^4}\frac{256 m_{\chi}^2m_T^2}{(q^2-m_a^2)^2}\right)\frac{|\vec{q}|^4}{2J+1}\frac{(2J+1)(J+1)}{J}\nonumber \\
&&\times \left(a_p(q^2)\langle S_p\rangle+a_n(q^2)\langle S_n\rangle\right)^2,
\end{eqnarray}
where $J$ is the nucleus spin, $\langle S_{p,n}\rangle$ is the average spin of protons and neutrons.
In the case of xenon targets, $\langle S_{p}\rangle \ll \langle S_{n}\rangle$ for $^{129}$Xe and $^{131}$Xe~\cite{Vietze:2014vsa}, we can make an approximate formula
\begin{eqnarray}
\overline{\left| {\cal M}_T\right|^2}
&\simeq&\frac{|\vec{q}|^4}{16} \left(\frac{\xi^2}{f^4}\frac{256 m_{\chi}^2m_T^2}{(q^2-m_a^2)^2}\right)\nonumber\\
&&\times a_n^2(q^2)\frac{4\pi}{(2J+1)}{\cal S}_n(q^2)\, ,
\end{eqnarray}
where
\begin{eqnarray}
{\cal S}_n(q^2)\equiv \frac{(2J+1)(J+1)}{\pi J}\langle S_n\rangle^2(q^2)\, ,
\end{eqnarray}
The structure factors ${\cal S}_n(q^2)$ depend on the momentum transfer and the numerical values are taken from Ref.~\cite{Vietze:2014vsa}.

In the lab frame, the differential cross section is expressed by
\begin{eqnarray}
d\sigma &=& \frac{1}{2m_T 2E_{\chi,i} v} \frac{d^3 p_{T,f}}{(2\pi)^3 E_{T,f}} \frac{d^3 p_{\chi,f}}{(2\pi)^3 E_{\chi,f}}\nonumber\\
&&\times (2\pi)^4 \delta^4 (p_{\chi,i}+p_{T,i}-p_{\chi,f}-p_{T,f}) \overline{ \vert \mathcal{M}_T\vert^2} \nonumber\\
&&= \frac{1}{64\pi^2 m_T E_{\chi,i} v} \frac{\vert\vec{p}_{\chi,f} \vert}{E_{T,f}}d\Omega \overline{ \vert \mathcal{M}_T\vert^2} \nonumber\\
&&=\frac{\vert\vec{p}_{\chi,f} \vert}{64\pi^2 m^2_T m_{\chi} v}  d\Omega \overline{ \vert \mathcal{M}_T\vert^2}, 
\label{eq:dsigmadEr}
\end{eqnarray}
where initial (final) momenta and energy are denoted by $p_i (p_f), ~E_i(E_f)$, respectively.
We have taken the approximation, $E_{\chi,i} \approx m_\chi,~E_{T,f}\approx m_T$.
The recoil energy is very small in the scattering process, so it is evaluated in the non-relativistic limit.
The recoil energy is 
\begin{align}
E_R &= \frac{ \vert \vec{q}\vert^2}{2m_T} = \frac{\vert \vec{p}_{\chi,i}\vert^2+\vert \vec{p}_{\chi,f}\vert^2 -2\vert \vec{p}_{\chi,i}\vert\vert \vec{p}_{\chi,f}\vert \cos\theta}{2m_T}\, .
\end{align}
We can obtain 
\begin{align}
dE_R &= - \frac{ \vert \vec{p}_{\chi,i}\vert\vert \vec{p}_{\chi,f}\vert  }{m_T} d\cos\theta \rightarrow d\Omega = 2\pi \frac{ m_T }{  \vert \vec{p}_{\chi,i}\vert\vert \vec{p}_{\chi,f}\vert  } dE_R. 
\label{eq:dOmegadEr}
\end{align}
Here we have used $d\Omega=-2\pi d\cos\theta$.
Using Eqs.~\eqref{eq:dsigmadEr} and \eqref{eq:dOmegadEr}, we can construct the differential scattering cross section that reads \cite{Fan:2010gt, Cirelli:2013ufw}
\begin{align}
\frac{d\sigma}{dE_R} &= \frac{1}{32\pi m^2_{\chi} m_T v^2}  \overline{ \vert \mathcal{M}_T\vert^2}.
\label{eq:diff_cross}
\end{align}
The differential cross section for the DM-nucleus scattering is thus given by
\begin{eqnarray}
\frac{d\sigma}{dE_R}
&=&\frac{2m_T}{(2J+1)}\frac{1}{v^2}\frac{\xi^2}{f^4}\frac{|\vec{q}|^4}{(q^2-m_a^2)^2}{\cal S}_n(q^2)\nonumber\\
&\simeq&\frac{2m_T}{(2J+1)}\frac{1}{v^2}\frac{\xi^2}{f^4}\frac{(2m_TE_R)^2}{(2m_TE_R+m_a^2)^2}{\cal S}_n(q^2)\, ,
\label{eq:diff_xsec_single}
\end{eqnarray}
where we have taken the non-relativistic limit $q^2\simeq -|\vec{q}|^2\simeq -2m_TE_R$ in the second line.
In the limit of the light mediator, $2m_T E_R \gg m^2_a$ , the differential cross section becomes
\begin{align}
\frac{d\sigma}{dE_R}=\frac{2m_T}{(2J+1)}\frac{1}{v^2}\frac{\xi^2}{f^4}{\cal S}_n(q^2)\, ,
\end{align}
so the $|\vec{q}|^4$ suppression from the effective operator in Eq.~\eqref{eq:eff_sd} disappears.
In this case, therefore, significant DM-nucleus scattering can be possible via the axion mediated processes.
In the limit of the heavy mediator corresponding to $q^2\to 0$, the differential cross section is
\begin{align}
\frac{d\sigma}{dE_R}\simeq\frac{2m_T}{(2J+1)}\frac{1}{v^2}\frac{\xi^2}{f^4}\frac{(2m_TE_R)^2}{m_a^4}{\cal S}_n(q^2).
\end{align}
The DM-nucleus scattering via the axion-mediated processes suffers the $q^4$ suppression.

For a given recoil energy, the differential event rate for DM scattering off the xenon target in the unit of cpd (counts per day) per kilogram per keV, is then
\begin{align}
\frac{dR}{dE_R} &= \frac{\rho_{\chi}}{m_{\chi}} n_T \int_{v_{\rm min}}^{v_{\rm esc}} v f(\vec{v}) \frac{d\sigma}{dE_R} d^3v,
\label{eq:diff_rate}
\end{align}
where $\rho_{\chi}=0.3$~${\rm GeV/cm^3}$ is the local DM energy density, $n_T$ is the number density of the target atoms.
Two natural isotopes of xenon with nonzero nuclear spin are
$^{129}{\rm Xe}$ (spin $1/2$) and $^{131}{\rm Xe}$ (spin $3/2$),
and their natural abundances of 26.4\% and 21.2\%, respectively \cite{natural_abundances, crc}. 
The remaining 52.4\% of xenon is insensitive to the spin-dependent interaction~\cite{XENON:2019rxp}.
The DM velocity in the galactic frame is described by the truncated Maxwell-Boltzmann distribution which corresponds to the isothermal sphere density profile:
\begin{eqnarray}
f(\vec{v}) d^3\vec{v} &=& \frac{ N }{ (v_{\rm mp} \sqrt{\pi})^3} e^{-v^2/v^2_{\rm mp}}d^3\vec{v},
\end{eqnarray}
where the most probable veolicty is given by $v_{\rm mp}=238$km/s \cite{v02016, Abuter:2021}, and the distribution is truncated at the escape velocity $v_{\rm esc}=544$ km/s \cite{v02016}.
The normalization factor $N$ is 
\begin{eqnarray}
	N&=& \frac{1}{{\rm erf}(v_{\rm esc}/v_{\rm mp}) -\frac{2}{\sqrt{\pi}} (v_{\rm esc}/v_{\rm mp}) {\rm exp}(-v^2_{\rm esc}/v_{\rm mp}^2) }.
\end{eqnarray}
to make the distribution satisfy the condition, 
\begin{eqnarray}
	\int_{v\leq v_{\rm esc}} d^3v f(\vec{v} )=1\, .
\end{eqnarray}

\begin{figure}[t]
\centering
\includegraphics[width=0.8\linewidth]{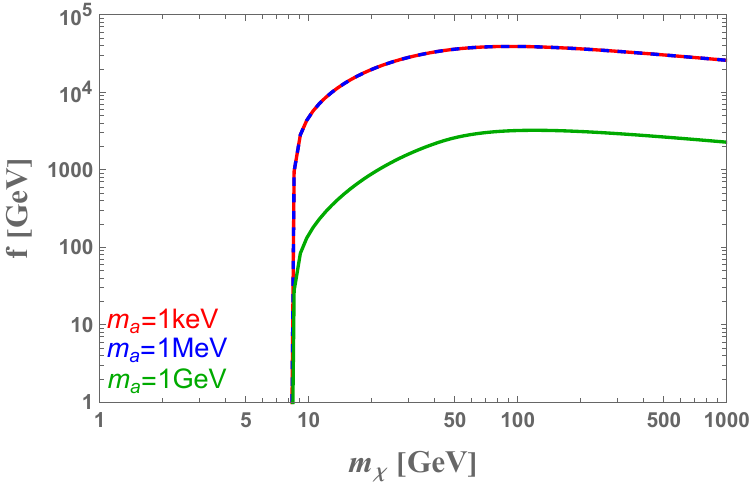} 
\vspace{0.5cm}
\caption{ Direct detection bound for the model with the elastic scattering case is shown on $(m_\chi, f)$-plane. Region below the lines are excluded by XENONnT. 
DM mass below $8$~GeV does not get any sensible bound due to the energy threshold. 
\label{DM-elastic-plot}} 
\end{figure}

The expected total event rate for DM scattering is 
\begin{eqnarray}
R &=& \int \epsilon(E_R)  \frac{dR}{dE_R} dE_R.
\end{eqnarray}
where $\epsilon(E_R)$ is detector efficiency.
For the recent result of the XENONnT experiment, an energy range above 10\% total efficiency lies between $3.3~{\rm keV_{NR}}$ to $60.5~{\rm keV_{NR}}$ and the amount of data is $1.1~\text{t}\cdot\text{yr}$~\cite{XENON:2023cxc}.
Assuming only the statistical uncertainties, the 90\% confidence level sensitivity is 2.3 events if there is a null DM scattering event.
We analyze the XENONnT bound on $(m_{\chi},f)$-plane for the simple model in Eq.~\eqref{eq:toy_model}.
In Fig.~\ref{DM-elastic-plot}, We show the XENONnT bound for three benchmark points with $m_a=1$~keV, $1$~MeV, and $1$~GeV.
For a small $m_a$, the cross section is nearly independent of the axion mass and consequently the explicit $E_R$ dependence is cancelled out,
so the XENONnT results have a rather strong sensitivity up to $f\sim10$~TeV region.
For a larger $m_a$, the cross section gets suppressed by the factor $E_R^2m_N^2/m_a^4$, so the sensitivity is weaker than the small mass cases.

\section{Multi-Component DM Model and Inelastic Processes}
\label{sec:2-comp}

\begin{figure}[t]
\centering
\includegraphics[width=0.65\linewidth]{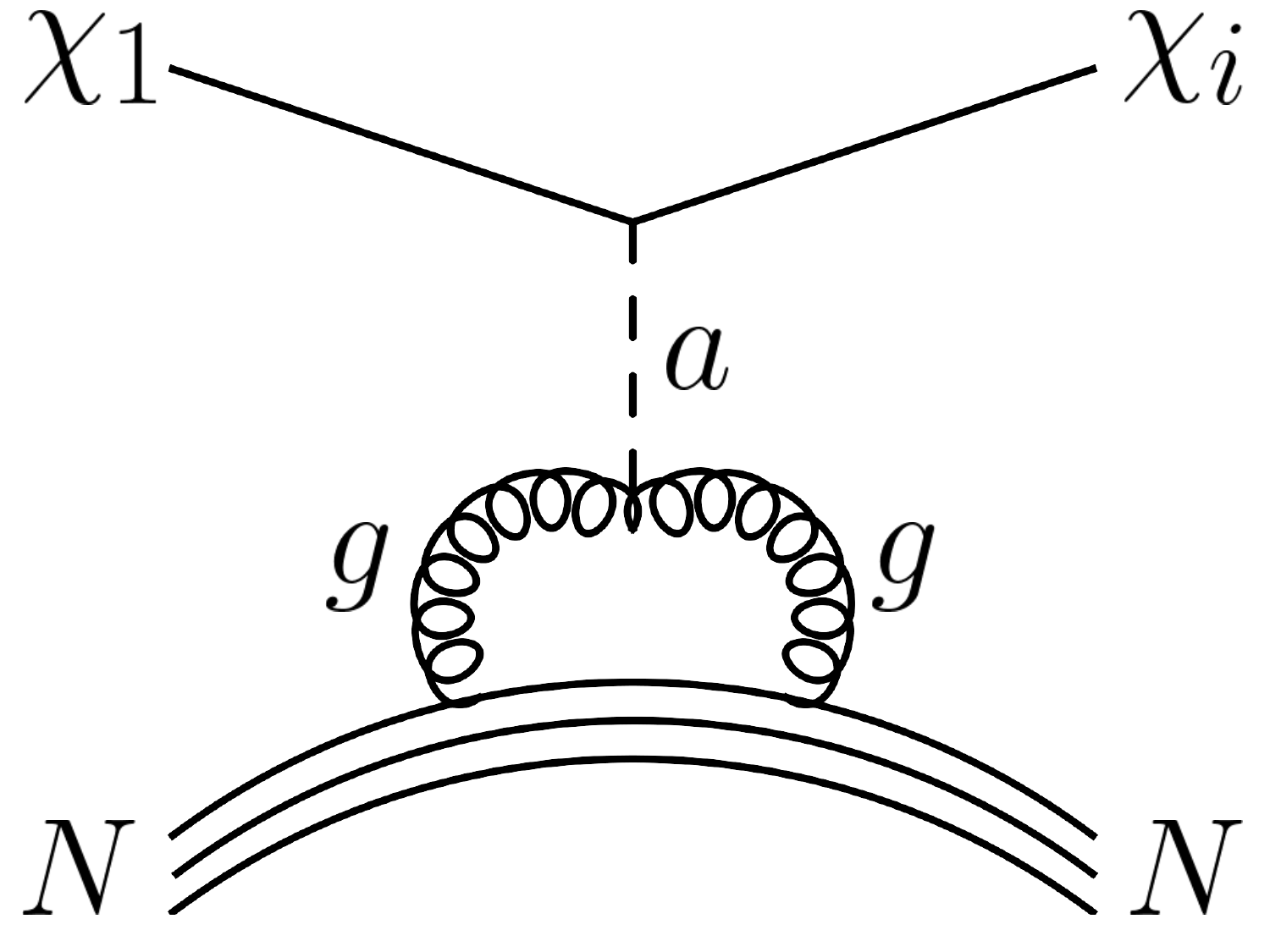} 
\vspace{0.5cm}
\caption{ Feynman diagram for both elastic and inelastic scattering process.} \label{inelastic_process}
\end{figure} 

We elevate the axion-mediated DM model to a multi-component DM model
in order to elucidate the impact of inelastic processes.
The Lagrangian includes the interactions which involve different DM components:
\begin{eqnarray} 
{\cal L}&=&{\cal L}_{\rm kin}+\frac12 m_{\chi_i}\bar{\chi_i}\chi_i - V(a)\nonumber \\
&&+\frac{g_s^2}{32\pi^2 f} aG_{\mu\nu}^b \tilde{G}^{b\mu\nu}
-\frac{\xi_{ij}}{f} \left(\partial_{\mu} a\right)\bar{\chi_i}\gamma^5\gamma^{\mu}\chi_j\, .\label{eq:multi-compt}
\end{eqnarray}
Here the DM masses are assumed to be diagonal while the coupling constant, $\xi_{ij}$ contains both diagonal and non-diagonal components.
The lightest component contributes to the dominant DM abundance, so is the initial state of the DM-nucleon scattering process.
The final state of the process, however, can be any component of $\chi_i$.
Therefore both the elastic and inelastic processes contribute to the DM-nucleon scattering as shown in Fig.~~\ref{inelastic_process}.

Let us consider a simple 2-component case where $\xi_{11}=\xi_{22}=\xi_{12}=\xi$. 
We assume that $\chi_1$ is the dominant DM in the universe, while $\chi_2$ is slightly heavier than $\chi_1$.
In this case, a sizable contribution to the DM detection is also made by the inelastic process, $\chi_1 N\to\chi_2N'$ if the DM particle has enough energy.
In the non-relativistic limit, the recoil energy of the scattering process is given by \cite{Bramante:2016rdh}
\begin{align}
E_R = \frac{\mu}{m_T} &\left[ \mu v^2 \cos^2\theta -\delta \pm \sqrt{ \mu v^2\cos^2\theta  } \sqrt{ \mu v^2\cos^2\theta -2\delta   }  \right],
\end{align}
where $m_T$ is the target nuclear mass, $\mu$ is the reduced mass between the DM particle and target nucleus, $\theta$ is the scattering angle in the lab frame, and $\delta=m_{\chi_2}-m_{\chi_1}$.
In this expression, we can see both the lower and upper bound on the recoil energy. 

\begin{figure}[t]
\centering
\includegraphics[width=0.85\linewidth]{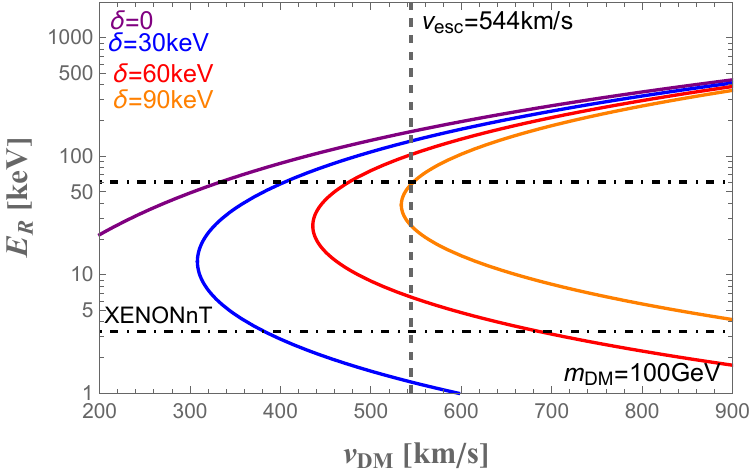} 
\vspace{0.5cm}
\caption{\label{velocity_VS_ER} The available range of recoil energies on a nuclear target, for a given DM mass. The dot-dashed black horizontal lines indicate the minimum and maximum recoil energy windows used by XENONnT collaboration \cite{XENON:2023cxc}. The gray vertical line is the DM escape velocity. 
} 
\end{figure}

The smallest value for the required DM velocity is given by
\begin{eqnarray}
v_{\rm min} = \frac{1}{\sqrt{2E_R m_T}} \left( \frac{E_R m_T}{\mu} +\delta \right)\, ,
\end{eqnarray}
and the recoil energy for $v_{\rm min}$ is given by
\begin{eqnarray}
E_R(v_{\rm min}) \simeq
\begin{cases}
\delta &(\text{for}~~m_\chi \gg m_T)\, , \\
 \frac{m_\chi}{m_T}\delta  &(\text{for}~~m_\chi \ll m_T)\, .
\end{cases}
\end{eqnarray}
In Fig.~\ref{velocity_VS_ER}, we show the available recoil energies on a xenon target depending on $\delta$ for $m_{\rm DM}=100$~GeV.
The available recoil energy region for the XENONnT experiment is between horizontal lines.
The vertical line corresponds to the escape velocity of DM in our galaxy. 
We can see that the phase space of the inelastic scattering shrinks and thus the event rate decreases as $\delta$ 
increases.
When $\delta$ is larger than $90$~keV, the phase space of the inelastic scattering is closed and only the elastic scattering process is involved.

From the Lagrangian in Eq.~\eqref{eq:multi-compt},
we can obtain the effective operators in a similar form as Eq.~\eqref{eq:DM-nuc_int}, 
but two Majorana fermions can be different:
\begin{eqnarray}
{\cal L}={\cal O}_{\chi_i N} (\bar{\chi}_i\gamma^5\chi_1)(\bar{N}\gamma^5N)\, ,
\label{eq:eff_lag_inel}
\end{eqnarray}
where
\begin{eqnarray}
{\cal O}_{\chi_i N}=-\left[\frac{\xi_{i1}}{f^2}\frac{2(m_{\chi_i}+m_{\chi_1})m_N}{q^2-m_a^2}\right]a_N(q^2)\, .
\label{eq:inel_eff}
\end{eqnarray}
For the realistic inelastic process, $\delta$ is much smaller than $m_1$ so we take $m_{\chi_i}\simeq m_{\chi_1}$ in Eq.~\eqref{eq:inel_eff}.
In the non-relativistic limit with small DM velocity, furthermore, we have the same form of the spin interactions as Eq.~\eqref{eq:eff_sd},
since the 3-momentum of incident DM particle and mass difference are much smaller than the DM mass.

The differential cross section is thus of the same form as Eq.~\eqref{eq:diff_xsec_single}:
\begin{eqnarray}
\frac{d\sigma_i}{dE_R}
&\simeq&\frac{2m_T}{(2J+1)}\frac{1}{v^2}\frac{\xi_{i1}^2}{f^4}\frac{(2m_TE_R)^2}{(2m_TE_R+m_a^2)^2}{\cal S}_n(q^2)\, ,
\label{eq:diff_xsec_multi}
\end{eqnarray}
where $\sigma_i$ corresponds to $\chi_1\to \chi_i$ transition during the DM-nucleus scattering process.
The differential event rate is also given by the same expression:
\begin{align}
\frac{dR_i}{dE_R} &= \frac{\rho_{\chi_1}}{m_{\chi_1}} n_T  \int_{v_{\rm min}}^{v_{\rm esc}} v f(\vec{v}) \frac{d\sigma_i}{dE_R} d^3v,
\label{eq:multi_diff_rate}
\end{align}
\begin{figure}[t]
\centering
\includegraphics[width=0.85\linewidth]{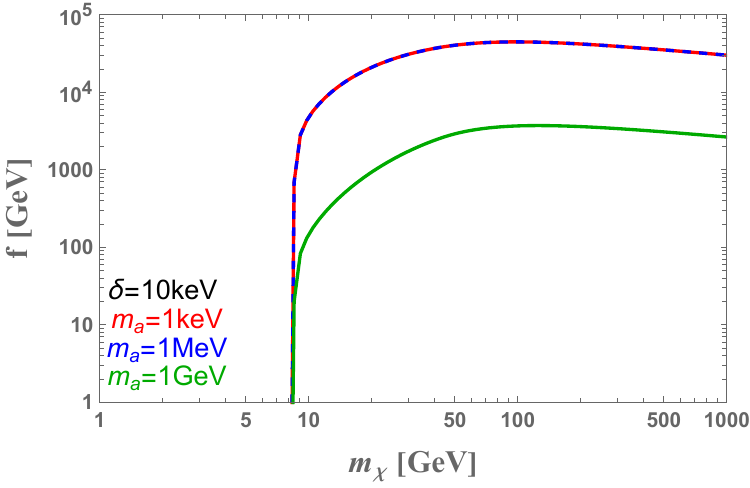} 
\includegraphics[width=0.85\linewidth]{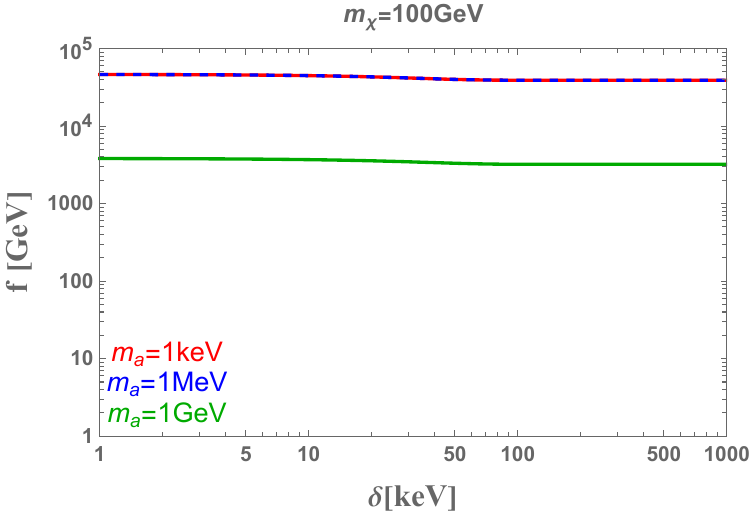} 
\vspace{0.5cm}
\caption{The XENONnT bound for the 2-component model is shown on $(m_\chi, f)$-plane in the upper panel and
in $(\delta, f)$-plane in the lower panel. Due to the energy threshold, DM mass below $8$~GeV is not constrained.
\label{DM-Inelastic-plot}} 
\end{figure}
%
The expected total event rate for DM is the sum of the elastic and inelastic processes:
\begin{eqnarray}
R =\sum_{i=1}^2  \int \epsilon(E_R) \frac{dR_i}{dE_R} dE_R\, .
\label{eq:tot_rate}
\end{eqnarray}
Each process implicitly depends on the mass difference $\delta$ in the $E_R$ and $v_{\rm min}$ in Eqs.~\eqref{eq:multi_diff_rate} and
\eqref{eq:tot_rate}, although the explicit $\delta$ dependence is negligible in the differential cross section in Eq.~\eqref{eq:diff_xsec_multi}.

In Fig.~\ref{DM-Inelastic-plot}, we show the XENONnT bound for the 2-component model.
The bound is similar to the single component model because the elastic and inelastic process are similar in the cross section.

\section{Supersymmetric Clockwork Axion Model}

\label{sec:susy_axion_model}

\subsection{Supersymmetric Clockwork Mechanism}

In this section, we briefly present the supersymmetric (SUSY) clockwork (CW) axion model and its dark matter sector 
as a well-motivated and well-organized model for axion-mediated dark matter which contains many fermion states.
We follow the basic construction in Ref.~\cite{Bae:2020hys} which utilizes a simple formulation in Ref.~\cite{Kaplan:2015fuy,Giudice:2016yja} for realizing the CW structure.
The model consists of $N+1$ superfield $\Phi_j$ containing the pseudo-Nambu-Goldstone Bosons (pNGBs) and their SUSY partners:
\begin{eqnarray}
\Phi_j=\frac{1}{\sqrt{2}}(\sigma_j+i\phi_j)+\sqrt{2}\theta\psi_j+\theta^2 F_j
\end{eqnarray}
where $\sigma_j$, $\phi_j$ and $\psi_j$ are the scalar, pseudoscalar and fermion components of the superfield.
The pseudoscalar $\phi_j$'s correspond to the pNGBs reflecting the shift symmetries originating from the spontaneously broken global U(1) symmetries. 

The effective K\"ahler potential and superpotential are given by
\begin{align}
K_{\rm eff}&=v_0^2\sum_{j=0}^N\left[\cosh\left(\frac{\Phi_j+\Phi_j^{\dagger}}{v_0}\right)+\xi\sinh\left(\frac{\Phi_j+\Phi_j^{\dagger}}{v_0}\right)\right]\, ,\label{eq:kahler}\\
W_{\rm eff}&=_{\Phi}v_0^2\sum_{j=0}^{N-1}\cosh\left(\frac{\Phi_j-c\Phi_{j+1}}{v_0}\right)\, ,
\end{align}
where $v_0$ is the mass scale for the U(1) symmetry breaking, $\xi$ is a dimensionless parameter, $m_{\Phi}$ is a mass parameter,
and $c$ is a natural number of order unity.
In the limit of $m_{\Phi}\to0$, the theory is invariant under the field transformation
\begin{eqnarray}
\phi_j\to \phi_j+\beta_j
\end{eqnarray}
where $\beta_j$'s are arbitrary real numbers.
Thus the $N+1$ shift symmetries are manifestly preserved.
For $m_{\Phi}\ne0$, $N$ shift symmetries become broken leaving one shift symmetry corresponding to
\begin{eqnarray}
\phi_j\to\phi_j+\beta/c^j\, ,
\end{eqnarray} 
where $\beta$ is an arbitrary real number.
This feature is clearly shown when the superpotential is expanded to the quadratic order:
\begin{eqnarray}
W_{\rm eff}&=&\frac12m_{\Phi}v_0^2\sum_{j=0}^{N-1}\left(\frac{\Phi_j-c\Phi_{j+1}}{v_0}\right)^2+\cdots\nonumber\\
&=&\frac12 m_{\Phi} \sum_{j,k=0}^N {\bf M}_{\text{CW} jk}\Phi_j\Phi_k+\cdots\, ,
\end{eqnarray}
where we have dropped the constant term.
The CW matrix $\bf M_{\rm CW}$ is given by
\begin{equation}
{\mathbf M}_{\rm CW}=
\begin{pmatrix}
1 & -c & 0 & \cdots & & 0 \\
-c & 1+c^2 & -c & \cdots & & 0 \\
0 & -c & 1+c^2 & \cdots & & 0 \\
\vdots & \vdots& \vdots& \ddots & & \vdots \\
& & & & 1+c^2 & -c \\
0 & 0 & 0 & \cdots & -c & c^2 
\end{pmatrix}.
\end{equation}
Hence the whole superfield is diagonalized by a single orthogonal matrix $\bf O$ so that
\begin{eqnarray}
\Phi_j={\bf{O}}_{jk}A_k
\end{eqnarray}
where one can call $A_k$ is the axion superfield.
The CW matrix is diagonalized by $\bf{O}$ 
\begin{eqnarray}
{\bf O}^T{\bf M}_{\rm CW}{\bf O} ={\rm diag}(\lambda_0,\cdots,\lambda_N)\, .
\end{eqnarray}
The eigenvalues and mixing matrix components are given by
\begin{eqnarray}
&&\lambda_0=0, \qquad \lambda_k =c^2+1-2c\cos\left(\frac{k\pi}{N+1}\right)\, , \label{eq:eigen_values}\\
&&{\mathbf O}_{j0}=\frac{{\cal N}_0}{c^j}, \nonumber\\
&&{\mathbf O}_{jk}={\cal N}_k\left[c\sin\frac{jk\pi}{N+1}-\sin\frac{(j+1)k\pi}{N+1}\right],
\label{eq:mix_mat}\\
&&\text{for}\quad j=0,\cdots,N;~~ k=1,\cdots,N \ ,\nonumber
\end{eqnarray}
where
\begin{eqnarray}
{\cal N}_0=\sqrt{\frac{c^2-1}{c^2-c^{-2N}}},\quad
{\cal N}_k=\sqrt{\frac{2}{(N+1)\lambda_k}}.
\end{eqnarray}
The zero-mode axion superfield $A_0$ is massless and the heavier modes $A_k$'s have masses $m_k=m_{\phi}\lambda_k$.

\subsection{Supersymmetry breaking contributions and mass spectra}

In the SUSY limit, the whole axion superfields have the same masses and interaction structure as those in the non-SUSY case shown in Ref.~\cite{Choi:2014rja,Choi:2015fiu,Kaplan:2015fuy,Higaki:2015jag,Giudice:2016yja}.
The axion superfields, however, undergo SUSY breaking effects which modify the mass spectra of scalars, pseudoscalars and fermions.
The superpotential terms acquire the SUSY breaking effect as follows:
\begin{eqnarray}
{\cal L}=\int d^2\theta(1+m_s\theta^2)W_{\rm eff}+\text{h.c.}
\end{eqnarray}
where one can extract the SUSY breaking effects in the scalar and pseudoscalar potentials,
\begin{eqnarray}
V_{\sigma}&=&-2m_{\rm sb}^2\sum_{j=0}^{N-1}\cosh\left(\frac{\sigma_j-c\sigma_{j+1}}{\sqrt{2}v_0}\right)\, ,\\
V_{\phi}&=&-2m_{\rm sb}^2\sum_{j=0}^{N-1}\cos\left(\frac{\phi_j-c\phi_{j+1}}{\sqrt{2}v_0}\right)\, ,
\end{eqnarray}
where the mass parameter is given by
\begin{eqnarray}
m_{\rm sb}^2\equiv m_{\Phi}|m_s|^2\cos\delta_s\, .
\end{eqnarray}
and $\delta_s$ is the phase of $m_s$.
In addition, the K\"ahler potential terms also acquire the SUSY breaking effects which contribute to the diagonal mass terms for scalars and fermions.
We assume that these contributions are universal for scalars and fermions. 

Including the SUSY-preserving and SUSY-breaking mass terms, the mass matrices are given by
\begin{eqnarray}
{\mathbf M}^2_{\phi}&=&m_{\Phi}^2{\mathbf M}^2_{\rm CW}
+m_{\rm sb}^2{\mathbf M}_{\rm CW},\\
{\mathbf M}^2_{\sigma}&=&m_{\Phi}^2{\mathbf M}^2_{\rm CW}
-m_{\rm sb}^2{\mathbf M}_{\rm CW}+\left(m_{\sigma}^{K}\right)^2{\mathbf I},\\
{\mathbf M}_{\psi}&=&m_{\Phi}{\mathbf M}_{\rm CW}+m_{\psi}^{K}{\mathbf I}.
\end{eqnarray}
where $(m_{\sigma}^K)^2$ and $m_{\psi}^K$ are the mass parameters generated by the SUSY breaking in the K\"ahler potential.
In this setup, an important thing is that all mass matrices are diagonalized by the {\it same} mixing matrix $\bf{O}$ as given in Eq.~\eqref{eq:mix_mat}.
The mass eigenstates for the pseudoscalars, scalars and fermions are respectively given by
\begin{eqnarray}
\phi_j&=&{\mathbf O}_{jk}a_k\, , \label{eq:axion_mx}\\
\sigma_j&=&{\mathbf O}_{jk}s_k\, ,\label{eq:sax_mx}\\
\psi_j&=&{\mathbf O}_{jk}\tilde{a}_k\, ,\label{eq:axn_mx}
\end{eqnarray}
and the mass eigenvalues are given by
\begin{eqnarray}
m_{a_k}^2&=&m_{\Phi}^2\lambda_k^2+m_{\rm sb}^2\lambda_k,\label{eq:axion_mass}\\
m_{s_k}^2&=&m_{\Phi}^2\lambda_k^2-m_{\rm sb}^2\lambda_k+(m_{\sigma}^K)^2,\label{eq:sax_mass}\\
m_{\tilde{a}_k}&=&m_{\Phi}\lambda_k+m_{\psi}^K\, .\label{eq:axn_mass}
\end{eqnarray}
It is noteworthy that the mass orders of the saxions and axinos may differ from that of the axions
because $m_{\rm sb}^2$ term in $m_{s_k}^2$ contains negative sign and $m_{\psi}^K$ can take either positive or negative value.
In our discussion, we always use the same index convention following the mixing matrices defined in Eqs.~\eqref{eq:axion_mx}, \eqref{eq:sax_mx} and \eqref{eq:axn_mx} regardless of the mass orders of saxions and axinos.
Therefore, for the axions, $a_0$ is the lightest mode while $s_0$ and $\tilde{a}$ can
be non-lightest modes in some parameter space. We will clarify our mass spectra of interest in the next section.

\subsection{Interactions}

The interaction of axion superfields with the SM sector is realized by introducing the coupling between
$\Phi_N$ and the SM gluons as
\begin{eqnarray}
{\cal L}=-\frac{g_s^2}{32\pi^2}\frac{1}{v_0}\int d^2\theta \Phi_N {\cal W}^{b\alpha}{\cal W}_{\alpha}^b+\text{h.c.}
\label{eq:PhiWW}
\end{eqnarray}
where ${\cal W}^b$ is the gluon superfield. 
It is also possible to introduce the coupling with the photon superfield, but we neglect it for simplicity.
The main feature of DM scattering does not alter even if the photon coupling is included.
Once the mass matrix is diagonalized, one can derive the axion superfield interaction from Eq.~\eqref{eq:PhiWW}:
\begin{align}
{\cal L}&=-\frac{g_s^2}{32\pi^2}\frac{1}{v_0}\int d^2\theta  {\cal W}^{b\alpha}{\cal W}_{\alpha}^b{\bf O}_{Nk}A_k+\text{h.c.}\nonumber\\
&=-\frac{g_s^2}{32\pi^2}\frac{1}{v_0}\int d^2\theta  {\cal W}^{b\alpha}{\cal W}_{\alpha}^b\nonumber\\
&\times\left[\frac{{\cal N}_0}{c^N} A_0-\sum_{k=1}^N(-1)^k{\cal N}_k c\sin\left(\frac{k\pi}{N+1}\right)A_k\right]+\text{h.c.}\, \label{eq:agg_int}
\end{align}
As in the non-SUSY CW axion model, the zero mode superfield suffer an exponential suppression $\sim 1/c^N$ in its interaction with the gluon
while the heavier modes experience a mild suppression $\sim1/N^{3/2}$.
For the component fields, the interaction terms are given by
\begin{eqnarray}
{\cal L}_{agg}=\frac{g_s^2{\bf O}_{Nk}}{32\sqrt{2}\pi^2 v_0} &&\left[a_kG^b_{\mu\nu}\tilde{G}^{b\mu\nu}+s_kG^b_{\mu\nu}G^{b\mu\nu} \right.\nonumber\\
&&\left.+\bar{\tilde{a}}_k\sigma^{\mu\nu}\gamma^5\tilde{g}^bG_{\mu\nu}^b\right]\, ,
\label{eq:agg_comp_int}
\end{eqnarray}
where $\tilde{g}$ is the gluino field.

The triple axion superfield interaction is also induced from the $\xi$-dependent term in the K\"ahler potential, Eq.~\eqref{eq:kahler}.
After diagonalizing the mass matrices, the interactions for component fields are given by
\begin{eqnarray}
{\cal L}_{aaa}&=&\frac{\xi}{\sqrt{2}v_0}\sum_{j}^N {\mathbf O}_{jn}{\mathbf O}_{jm}{\mathbf O}_{jl}\nonumber\\
&&\times\left[s_n(\partial_{\mu} a_m)(\partial^{\mu}a_l)+s_n(\partial_{\mu}s_m)(\partial^{\mu}s_l)\right.\nonumber\\
&&\left.+is_n\bar{\tilde{a}}_m\gamma^{\mu}\partial_{\mu}\tilde{a}_l
-(\partial_{\mu}a_n)\bar{\tilde{a}}_m\gamma^5\gamma^{\mu}\tilde{a}_l\right].
\label{eq:aaa_int}
\end{eqnarray}
Note that the whole interactions must sum over all $n$, $m$, $l$.

As we will see, these two couplings, axion-gluon-gluon term in Eq.~\eqref{eq:agg_comp_int} and axion-axino-axino in Eq.~\eqref{eq:aaa_int} generate the axion-mediated scattering process between the axino DM and nucleon.

\section{Axion-mediated dark matter in the SUSY CW axion model}
\label{sec:susy_axion_result}

\subsection{Mass spectra for axino dark matter}

We consider the axino DM in the SUSY CW axion model since it realizes a model of axion-mediated DM and inelastic scattering processes.
The model contains complex particle spectrum of the SUSY partners of the SM particles and axions.
For clear analyses, we assume that all SM partners and saxions are much heavier than the axions and axinos,
so the DM scattering processes are dominantly mediated by the axions.

The axion mass ordering in Eq.~\eqref{eq:axion_mass} is the same as that of the non-SUSY case since it is solely determined by
the CW matrix.
The axino mass ordering in Eq.~\eqref{eq:axn_mass} can, on the other hand, differ from that of the axions
while the interaction structure in Eq.~\eqref{eq:agg_comp_int} and \eqref{eq:aaa_int} does not alter.
In this respect, we consider two representative mass orderings: 1) $m_{\phi}$ and $m_{\psi}^K$ have the same sign (normal ordering) and
2) $m_{\psi}^K$ is large negative compared to $m_{\phi}$ so that $\tilde{a}_N$ becomes the lightest mode and $\tilde{a}_0$ is the heaviest mode (inverted ordering). The axino mass ordering is summarized in Table~\ref{tab:ordering}.
\begin{table}
\begin{tabular}{|c|c|c|}
\hline
ordering & normal & inverted \\
\hline
lightest  & $\tilde{a}_0$  & $\tilde{a}_N$  \\
(mass) & ($m_{\psi}^K$) & ($|m_{\psi}^K|-m_{\Phi}\lambda_N$) \\
\hline
heaviest & $\tilde{a}_N$  & $\tilde{a}_0$ \\
(mass) & ($m_{\psi}^K+m_{\Phi}\lambda_N$) & ($|m_{\psi}^K|$) \\
\hline
\end{tabular}
\caption{The axino masses for the normal and inverted orderings are shown. The $m_{\Phi}$ is assumed to be positive. 
The axino masses for the inverted ordering are made to be positive by fermion field redefinition.
\label{tab:ordering}}
\end{table}
We also stress that $m_{\Phi}$ controls the mass differences between the axino states while $m_{\Psi}^K$ determines the overall mass 
scale of the axinos.
The mass scale of non-zero mode axions is determined by $m_{\Phi}$ and $m_{\rm sb}$.

\subsection{Differential cross section for the scattering process}
One can easily generalize the effective Lagrangian for the axino (in)elastic scattering off the nuclei, $\tilde{a}_iN\to\tilde{a}_jN'$ 
by properly multiplying the CW mixing matrix components in Eqs.~\eqref{eq:agg_comp_int} and \eqref{eq:aaa_int} to that in Eq.~\eqref{eq:eff_lag_inel}, {\it i.e.},
\begin{eqnarray}
{\cal L}={\cal O}_{\tilde{a}_j\tilde{a}_i N N} (\bar{\chi}_j\gamma^5\chi_i)(\bar{N}\gamma^5N)\, ,
\label{eq:eff_lag_cw}
\end{eqnarray}
where
\begin{eqnarray}
{\cal O}_{\tilde{a}_j\tilde{a}_i N N}=-\frac{\xi}{f_a^2}   
\left[\sum_{k,l}^N 
\frac{{\mathbf O}_{Nl}{\mathbf O}_{kl}{\mathbf O}_{kj}{\mathbf O}_{ki}}{q^2 -m^2_{a_l}}\right]\nonumber\\
\times\left[2(m_{\tilde{a}_j}+m_{\tilde{a}_i})m_N\right]a_N(q^2)\, .
\label{eq:inel_eff_cw}
\end{eqnarray}
Here we identify $f_a=\sqrt{2}v_0$.
The initial state can be either $i=0$ for the normal ordering or $i=N$ for the inverted ordering.
The final state $j$ can be any number from $0$ to $N$ regardless of the mass ordering.
Consequently, the differential cross section for $\tilde{a}_iN\to\tilde{a}_jN'$ process is given by
\begin{eqnarray}
&&\frac{d\sigma}{dE_R}(\tilde{a}_iN\to\tilde{a}_jN')\nonumber\\
&&\simeq\frac{2m_T}{(2J+1)}\frac{1}{v^2}
\frac{\xi^2}{f_a^4}(2m_TE_R)^2 \nonumber\\  
&&\times\left|\sum_{k,l}^N 
\frac{{\mathbf O}_{Nl}{\mathbf O}_{kl}{\mathbf O}_{kj}{\mathbf O}_{ki}}{(2m_TE_R+m_{a_l}^2)}\right|^2
{\cal S}_n(q^2)\, .
\label{eq:diff_xsec_cw}
\end{eqnarray}
For the case of light axions, $m_{a_l}^2\ll m_TE_R$, one can simply neglect $m_{a_l}^2$ in the denominator in Eq.~\eqref{eq:diff_xsec_cw}.
The mixing matrix factors are further simplified to be
\begin{eqnarray}
 \sum_{k,l}^N {\mathbf O}_{Nl}{\mathbf O}_{kl}{\mathbf O}_{kj}{\mathbf O}_{ki}={\mathbf O}_{Nj}{\mathbf O}_{Ni}\, .
\end{eqnarray}
The differential cross section Eq.~\eqref{eq:diff_xsec_cw} becomes
\begin{eqnarray}
&&\frac{d\sigma}{dE_R}(\tilde{a}_iN\to\tilde{a}_jN')\nonumber\\
&&\simeq\frac{2m_T}{(2J+1)}\frac{1}{v^2}
\frac{\xi^2}{f_a^4}\left|{\mathbf O}_{Nj}{\mathbf O}_{Ni}\right|^2
{\cal S}_n(q^2)\, .
\end{eqnarray}
If we neglect the dependence of phase space of final state particles and mass difference $\delta$, 
we can use the total differential cross section for the light axion case as
\begin{eqnarray}
\left(\frac{d\sigma_{\rm tot}}{d E_R}\right)_{\rm la}&\simeq&\sum_j \frac{d\sigma}{dE_R}(\tilde{a}_iN\to \tilde{a}_j N') \nonumber\\
&\simeq&\frac{2m_T}{(2J+1)}\frac{1}{v^2}
\frac{\xi^2}{f_a^4}{\cal S}_n(q^2)
\sum_j \left|{\mathbf O}_{Nj}{\mathbf O}_{Ni}\right|^2 \nonumber\\
&=&\frac{2m_T}{(2J+1)}\frac{1}{v^2}
\frac{\xi^2}{f_a^4}{\cal S}_n(q^2)
\left|{\mathbf O}_{Ni}\right|^2
\label{eq:diff_cross_massless_axion}
\end{eqnarray}
where we have used the orthogonality, $\sum_j |\mathbf{O}_{Nj}|^2=1$ in the last line.
As shown in Eq.~\eqref{eq:diff_cross_massless_axion}, the total differential cross section depends on $|\mathbf{O}_{Ni}|^2$ in the limit of
light axions.
Because of this feature, the normal axino mass ordering case ($i=0$) suffers from huge suppression, $\sim 1/c^{2N}$ while the inverted ordering case do not get such a suppression factor.

For the case of heavy axions, $m_{a_l}^2\gg m_TE_R$ (for $l>0$), we can proceed the summation part in Eq.~\eqref{eq:diff_xsec_cw} as the following,
\begin{eqnarray}
{\cal P}_{Nk}&\equiv&\sum_{l=0}^N {\mathbf O}_{Nl}{\mathbf O}_{kl}
\frac{1}{2m_T E_R + m^2_{a_l}}\nonumber\\
&\simeq& \frac{{\cal N}_0^2}{c^{N+k}} \frac{1}{2m_TE_R}+
\sum_{l=1}^N{\mathbf O}_{Nl}{\mathbf O}_{kl}\frac{1}{m_{a_l}^2}\, 
\label{eq:factor_heavy_axion}
\end{eqnarray}
where the first term in the second line comes from the zero-mode axion contribution.
For realizing the QCD axion, we construct a model with $c^N\sim 10^6$ and $f_a\gtrsim\text{TeV}$.
The typical recoil energy is around tens of keV, so we expect that the zero mode term dominates for $m_{a_l}\gtrsim 100~\text{MeV}$.
We will clearly see this feature in the numerical analyses shown in the following subsection. 
In the limit of small mass difference $\delta$, the total differential cross section for the heavy axion case is given by
\begin{eqnarray}
\left(\frac{d\sigma_{\rm tot}}{d E_R}\right)_{\rm ha}
&\simeq&\frac{2m_T}{(2J+1)}\frac{1}{v^2}
\frac{\xi^2}{f_a^4}(2m_TE_R)^2 \nonumber\\  
&&\times\sum_j\left|\sum_{k}^N 
{\mathbf O}_{kj}{\mathbf O}_{ki}{\cal P}_{Nk} \right|^2
{\cal S}_n(q^2)\, .
\label{eq:tot_sigma_ha}
\end{eqnarray}

\subsection{Numerical analyses}

\begin{figure}[t]
\centering
\includegraphics[width=0.85\linewidth]{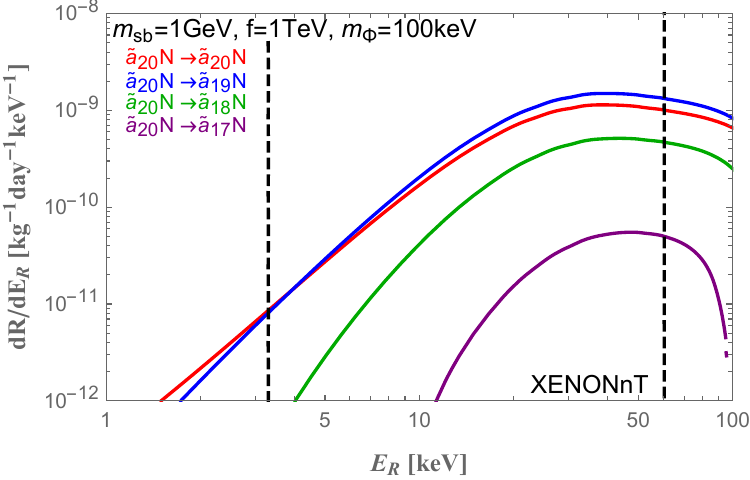} 
\vspace{0.5cm}
\caption{  \label{xenon:dRdER} The differential event rate are shown for the inverted ordering case. We take $m_{\chi_{1}}=100$~GeV, $m_{\rm sb}=1$~GeV, $f_a=1$~TeV and $m_\Phi=100$~keV. 
Here we can see three inelastic scattering processes.
}
\end{figure} 

\begin{figure}[t]
\centering
\includegraphics[width=0.9\linewidth]{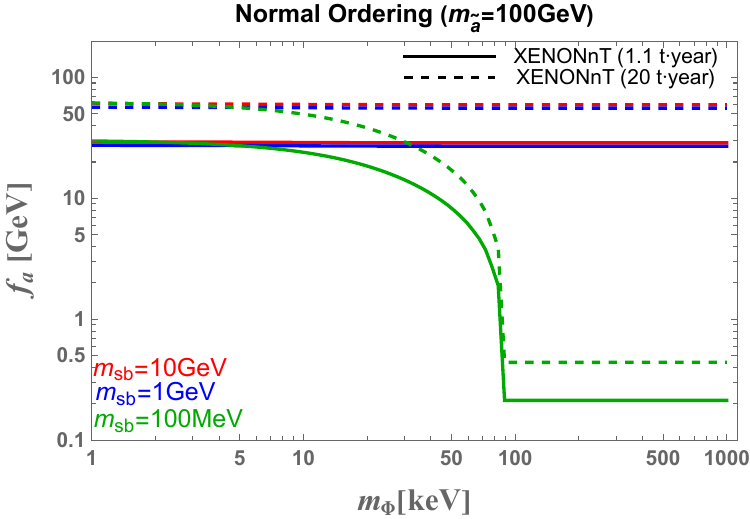} 
\includegraphics[width=0.9\linewidth]{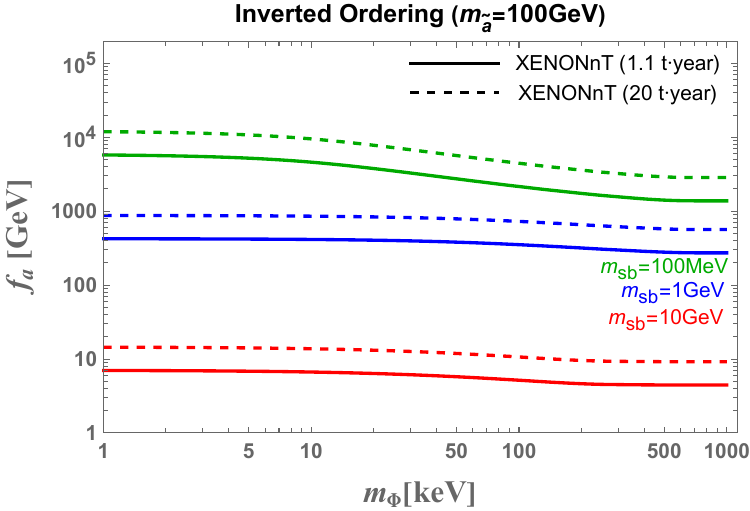} 
\vspace{0.5cm}
\caption{ \label{event:XENONnT} The XENONnT bounds are shown for the normal (uppper) and inverted (lower) mass ordering. The solid (dashed) lines represent current (future) XENONnT experimental bounds. We have taken the DM mass $100$~GeV for both cases and used $m_{\rm sb}= 10 (\text{Red}),~1(\text{Blue}),~0.1(\text{Green})$ GeV, respectively.}
\end{figure} 

For the numerical analyses, we consider a benchmark scenario with $\xi=1$, $c=2$ and $N=20$,
so we achieve a large hierarchy $c^N\sim10^6$.
In this case, the QCD axion is realized by the zero-mode axion ($a_0$) when $f_a\gtrsim1$~TeV and we have 21 axions and 21 axinos which are involved in the scattering processes.

In Fig.~\ref{xenon:dRdER}, we show the differential event rates, $dR/dE_R$, of the elastic and inelastic processes in the inverted ordering case.
Here we take $m_{\rm sb}=1$~GeV, $f_a=1$~TeV and $m_\Phi=100$~keV
where only three inelastic scattering processes appear.
When $m_{\Phi} \leq \mathcal{O}(10)$~keV, all inelastic scattering processes are involved in the DM-nucleon scattering.
In the inverted ordering case, at least one inelastic scattering process is included in the XENONnT analyses for $m_\Phi \lsim 700$~keV.
In the normal ordering case, on the other hand, at least one inelastic process is involved for $m_\Phi \lsim 100$~keV,
so we can see more inelastic processes for fairly small values of $m_{\Phi}$.
The difference between the normal and inverted ordering cases originates from the intrinsic structure of CW spectrum where the zero-indexed states ($a_0$, $s_0$, $\tilde{a}_0$) are relatively isolated from the other states.
In the inverted ordering case, therefore, the lightest axino ($\tilde{a}_{20}$) can be more easily excited to the heavier modes ($\tilde{a}_{19}$, $\tilde{a}_{18}$, $\cdots$) than it is in the normal ordering case. 

As we have seen in Fig.~\ref{velocity_VS_ER}, the available range of recoil energy shrinks as the mass difference increases.
This feature also appears in Fig.~\ref{xenon:dRdER}.
The full $E_R$ region of XENONnT is relevant for the elastic processes ($\tilde{a}_{20}N\to\tilde{a}_{20}N$) and two inelastic process ($\tilde{a}_{20}N\to\tilde{a}_{19}N$ and $\tilde{a}_{20}N\to\tilde{a}_{18}N$) while one inelastic process ($\tilde{a}_{20}N\to\tilde{a}_{17}N$) is not accessible for $E_R\lesssim9$~keV.

In Fig.~\ref{event:XENONnT}, we show the XENONnT bound on $(m_\Phi,f_a)$-plane.
We take the lightest axino mass $100$~GeV and three different $m_{\rm sb}= 100$~MeV, 1~GeV, 10~GeV.
The solid (dashed) lines represent the current (future) bounds from the XENONnT experiment. 
In order to derive the bounds, we use the same parameters as we did in Sec.~\ref{sec:single} and \ref{sec:2-comp}: recoil energy range above 10\% total efficiency in range of $3.3~{\rm keV_{NR}}$ to $60.5~{\rm keV_{NR}}$ and 2.3 events expected in the 90\% confidence level in the null hyperthesis.
We use $1.1$~t$\cdot$year for the current bound and $20$~t$\cdot$year for the future prospects.

In the case of the normal axino mass ordering, the XENONnT bounds are of the same order for the light axions ($m_{\rm sb}=100$~MeV) and heavy axions ($m_{\rm sb}=1$~GeV and $10$~GeV).
For the light axion case ($m_{\rm sb}=100$~MeV), the differential cross section is close to the approximate formula in Eq.~\eqref{eq:diff_cross_massless_axion},
and the bound for $f_a$ reaches around $30$~GeV with the current data and $60$~GeV with the future prospects.
As $m_{\Phi}$ increases, less inelastic processes are accessible and thus the total event rate decreases.
For $m_{\Phi}\gtrsim100$~keV, no inelastic processes are viable in the $E_R$ range of XENONnT,
so the only elastic scattering can survive  and the total cross section undergoes an additional $|{\cal O}_{N0}|^2\sim 1/c^{2N}$ suppression which is shown in the second line of Eq.~\eqref{eq:diff_cross_massless_axion}.
This suppression is transferred to the sensitivity to the $f_a$ bound by the factor $(1/c^{2N})^{1/4}\sim 10^{-3}$.
For a very small $m_{\rm sb}$, {\it e.g.}, $1$~MeV, the suppression becomes ${\cal O}(10^{-3})$.
As matter of fact,
for $m_{\rm sb}=100$~MeV, the bound receives the suppression of ${\cal O}(10^{-2})$ because $2m_T E_R$ is not much larger
than $m_{a_l}^2$,
so the XENONnT bound reaches only to around $300$~MeV with the current data and $600$~MeV with the future prospects.
This small values actually do not make any sensible bounds for the model because the effective theory approach for the SUSY CW axion model breaks down at this scale. 

For the heavy axion cases ($m_{\rm sb}=1$~GeV and $10$~GeV), 
the total scattering process follows the approximate relation in Eq.~\eqref{eq:tot_sigma_ha}.
Since we consider cases with small $m_{\Phi}$, the axion masses are determined dominantly 
by the $m_{\rm sb}$-dependent term in Eq.~\eqref{eq:axion_mass}.
The eigenvalues $\lambda_k$ in Eq.~\eqref{eq:eigen_values} varies in range from $(c-1)^2$ to $(c+1)^2$, which are of order unity.
We can take a crude simplification for the second term of Eq.~\eqref{eq:factor_heavy_axion}:
\begin{eqnarray}
\sum_{l=1}^N{\mathbf O}_{Nl}{\mathbf O}_{kl}\frac{1}{m_{a_l}^2}
\simeq \frac{1}{m_{\rm sb}^2}\sum_{l=0}^{N} {\mathbf O}_{Nl}{\mathbf O}_{kl}
\simeq \frac{\delta_{Nk}}{m_{\rm sb}^2}\, .
\end{eqnarray}
The total cross section becomes
\begin{eqnarray}
\left(\frac{d\sigma_{\rm tot}}{d E_R}\right)_{\rm ha}
&\simeq&\frac{2m_T}{(2J+1)}\frac{1}{v^2}
\frac{\xi^2}{f_a^4}(2m_TE_R)^2 {\cal S}_n(q^2)\nonumber\\  
&&\times\sum_j\left|\frac{{\cal N}_0^2}{c^{N}} \frac{1}{2m_TE_R}\sum_{k}^N 
\frac{{\mathbf O}_{kj}{\mathbf O}_{ki}}{c^{k}} \right. \nonumber\\
&&\left.+\frac{1}{m_{\rm sb}^2}{\mathbf O}_{Ni}{\mathbf O}_{Nj} \right|^2
\, .
\label{eq:heavy_axion_approx}
\end{eqnarray}
Here the sum of $j$ must be conducted for viable processes.
For the normal mass ordering ($i=0$), the first term in Eq.~\eqref{eq:heavy_axion_approx}
is dominant. 
Since the sum of $k$ is not much smaller than unity for $i=0$, Eq.~\eqref{eq:heavy_axion_approx} is
almost the same as the light axion case described by Eq.~\eqref{eq:diff_cross_massless_axion}.
This feature clearly appears in the upper panel of Fig.~\ref{event:XENONnT}.
In the heavy axion cases,
the dominant scattering processes is mediated by the zero-mode axion, {\it i.e.} QCD axion,
so the XENONnT sensitivity   
is directly interpreted as the bound for the QCD axion as a mediator ($f_{\rm QCD}\sim f_a\times c^{2N}\gtrsim 3\times10^7$~GeV).
The bound shown here is thus far weaker than the astrophysical ones ($f_{\rm QCD}\gtrsim10^9$~GeV).

In the case of the inverted axino mass ordering, the XENONnT bounds for the light axion case ($m_{\rm sb}=100$~MeV)
follows Eq.~\eqref{eq:diff_cross_massless_axion} as in the case of the normal axino mass ordering.
In this case, however, there is no exponential suppression in $\mathbf{O}_{Ni}$ for $i=N$,
so we achieve a better sensitivity than that in the normal ordering case by a factor of ${\cal O}(10^2)$.
For $m_{\rm sb}=1$~GeV, the total cross section is dominantly determined by the second term of Eq.~\eqref{eq:heavy_axion_approx}, and it gives weaker sensitivity.
For $m_{\rm sb}=10$~GeV, the total cross section is dominantly determined by the first term of Eq.~\eqref{eq:heavy_axion_approx}.
Differently from the normal mass ordering case, the sum of $k$ is highly suppressed for $i=N$.
Hence the bound is much weaker than that for the light axion case by a factor of ${\cal O}(10^{-3})$.

\subsection{Other constraints}
Before closing this section, it is worth mentioning the possible constraints on the benchmark scenarios discussed here.

\subsubsection{Constraints on axions}

Since there are many axion states, a number of constraints may be applicable from the collider~\cite{Mimasu:2014nea, Bauer:2017ris}, flavor probes~\cite{Bauer:2021mvw}, beam dump experiments~\cite{Dobrich:2015jyk}, and cosmology/astrophysics~\cite{Depta:2020wmr,Balazs:2022tjl}.
Some constraints stem from the axion-photon interaction.
This is also applicable since we need to introduce the axion-photon coupling for the axions to decay although the coupling itself does not alter our analyses given in this section.
Since we need to require $f_a\gtrsim1$~TeV to incorporate the QCD axion, the collider bounds are almost irrelevant while some beam dump experiments
produce marginal constraints for $m_{\rm sb}\lesssim {\cal O}(100)$~MeV (corresponding to $m_{a_l}\lesssim {\cal O}(100)$~MeV for $l\ne0$).
Meanwhile, the flavor  bounds are applicable if the axions have flavor-changing couplings with quarks and leptons 
which are not in our interest.
The most stringent constraints, indeed, are derived from the cosmology/astrophysics.
Similar to those from the beam dump experiments,
there are marginal constraints for $m_{\rm sb}\lesssim 100$~MeV and $f_a\lesssim 1$~TeV.

\subsubsection{Axino relic abundance and indirect detection}


In the parameter space of our interest, the axino DM is produced via the gluino-mediated process, axion/saxion-mediated process, and saxion/axino decay.
If the saxions and axions are much heavier than the axinos (more specifically, $m_{s/a}\gg\sqrt{m_{\tilde{g}}m_{\tilde{a}}}$), the dominant production is the gluino-mediated process.
In the case of axion-mediated DM direct detection, however, axions are not much heavier than the axinos.
Therefore in this case the the ratio between the gluino-mediated process and axion-mediated process is given by~\cite{Bae:2020hys}
\begin{eqnarray}
R\equiv \frac{\text{gluino-mediated}}{\text{axion-mediated}}=\frac{\alpha_s^2}{8\pi^2\xi^2}\frac{(s-m_{a_l}^2)^2}{m_{\tilde{g}}^2(m_{\tilde{a}_n}+m_{\tilde{a}_m})^2}
\end{eqnarray}
where $\alpha_s=g_s^2/(4\pi)$ and $s$ is the square of total energy of the process.
For the sensible XENONnT bound, the axions are ${\cal O}(100)$~MeV - ${\cal O}(1)$~GeV,\footnote{The saxion-mediated process can be also relevant. For a simple discussion, we assume that the saxions are much heavier than the axions.}
so the gluino-mediated process is dominant at a high temperature while the axion-mediated process is dominant at a low temperature.
In the low $f_a$ region, the interactions strength of the axino DM is not much different from that of typical WIMP candidates,
so the DM relic abundance is determined by the freeze-out process for the reheat temperature $T_R\gtrsim{\cal O}(10)$~GeV.
The relic abundance is determined by the axion-mediated processes, $gg\to a_{l}^*\to \tilde{a}\tilde{a}$, and thus the relic abundance is given by~\cite{Dror:2023fyd}
\begin{eqnarray}
\Omega_{\tilde{a}}\sim 4\times \Omega_{\rm DM}\left(\frac{f_a}{\text{TeV}}\right)^4\left(\frac{100\text{ GeV}}{m_{\tilde{a}}}\right)^2
\end{eqnarray}
It is worth noting that this relation is derived for the single axion case, 
and is applicable to the CW axion case if all non-zero mode axions are not very heavy
due to the orthogonality of the CW matrix in Eq.~\eqref{eq:mix_mat}.
The right relic abundance is obtained for $f_a\sim$~TeV.
For larger $f_a$, the axino DM from the freeze-out is overabundant, so we need a late time entropy production or low reheat temperature
which leads to the freeze-in production.

Some indirect detection constraints can be applicable if the axino DM has the electroweak interactions~\cite{Allen:2024ndv}.
The bound is $f_a\gtrsim{\cal O}(100)$~GeV for $m_{\tilde{a}}=100$~GeV, so it is comparable to the bound we obtained from
XENONnT for $m_{\rm sb}={\cal O}(1)$~GeV or weaker the that for $m_{\rm sb}={\cal O}(100)$~MeV in the case of the inverted mass ordering.


\section{Conclusions}

\label{sec:conclusion}

The DM-nucleus scattering process relies on how DM particles interact with the SM particles.
The massive part of DM study is based on the assumption that DM has interactions
of order the weak scale.
In this respect, it has been studied that the DM-nucleus scattering process is mediated by
the weak gauge bosons, Higgs boson, neutrinos, or new gauge bosons with weak scale couplings.
In this paper, we consider another possibility where the DM-nucleus scattering process is
mediated by the axions.
If the axions couple to the gluons via the QCD anomaly coupling, {\it i.e.}, $aG\tilde{G}$, the coupling induces the CP-odd gluonic current
and determines the gluon form factor of the nucleon.
Hence the DM-nucleus scattering process is made by exchanging the axion states as shown in Fig.~\ref{DM-elastic-fig}.

The simplest model can be built with the QCD axion because it is a necessary ingredient to solve the strong CP problem.
The QCD axion, however, has too small interactions with the gluons to be detected in the current experiments.
Instead, we can introduce a non-QCD axion which can mediate DM-nucleus scattering process.
We consider a simple model with a non-QCD axion coupled to a Majorana DM particle in Sec.~\ref{sec:single}.
We study how the CP-odd gluonic current contributes to the DM-nucleus scattering and show that it can make relevant constraints from the current XENONnT results as shown in Fig.~\ref{DM-elastic-plot}.

We extend our scope into a multi-component DM sector resulting in the axion-mediated inelastic scattering in Sec.~\ref{sec:2-comp}.
The DM inelastic scattering can be realized in some $E_R$ range of the XENONnT experiment for a given mass difference $\delta$.
The leading order calculation 
does not depend on the $\delta$, so we can simply obtain the differential cross section for both the elastic and inelastic processes in Eq.~\eqref{eq:diff_xsec_multi}.
In Fig.~\ref{DM-Inelastic-plot}, we show the XENONnT bounds for the 2-component model, and the bounds are similar to the single component case.

In Sec.~\ref{sec:susy_axion_model},
we consider the SUSY CW axion model to realize the QCD axion and the
axion-mediated DM which contain inelastic scattering processes.
In this model, the lightest axino can be a DM component and produce the DM-nucleus scattering processes mediated by the axions.
From the study on the CP-odd gluonic current and kinematics for the inelastic scattering,
we calculate the possible elastic and inelastic processes and show the current and future bounds for the model from the XENONnT experiments in Sec.~\ref{sec:susy_axion_result}.
The current XENONnT data already constrain some region of the parameter space for the inverted axino mass ordering case
while more data are required for achieving the relevant constraints for the normal ordering case.

For conclusions, the axion-mediated process can provide substantial contributions to the DM-nucleus scattering for both the elastic and inelastic cases.
The realistic scenarios can be accomplished from the SUSY CW axion model
and and the XENONnT results can reach significant sensitivity for some parameter region of the model.

\acknowledgments
This work was supported by National Research Foundation of Korea (NRF) Research Grant NRF-2022R1A5A1030700 (KJB) and 2019R1A2C3005009, 2022R1A2C2003567, RS-2024-00341419 (JK).

\bibliographystyle{utphys}
\bibliography{bib_MIDM}

\end{document}